\documentstyle[11pt,IAUS212,twoside,epsf]{article}

\def\edcomment#1{\iffalse\marginpar{\raggedright\sl#1\/}\else\relax\fi}
\reversemarginpar

\def\HeI{He\,{\sc i}}

\def\Mdot{\hbox{$\dot {M}$}}

\def\Lsun{\hbox{\it L$_\odot$}}

\def\Msun{\hbox{\it M$_\odot$}}
\def\Minit{\hbox{\it M$_{\rm initial}$}}
\def\Msunyr{\hbox{\it M$_\odot\,$yr$^{-1}$}}
\def\Myr{\hbox{\it Myr}}
\def\Gyr{\hbox{\it Gyr}}

\def\Vinf{\hbox{$v_\infty$}}
\def\kms{\hbox{km$\,$s$^{-1}$}}

\def\AK{\hbox{\it A$_{\rm K}$}}

\def\simgr{\mathrel{\hbox{\rlap{\hbox{\lower4pt\hbox{$\sim$}}}\hbox{$>$}}}}
\def\simls{\mathrel{\hbox{\rlap{\hbox{\lower4pt\hbox{$\sim$}}}\hbox{$<$}}}}
\def\mnk{\hbox{\it m$_{\rm F205W}$}}
\def\mnh{\hbox{\it m$_{\rm F160W}$}}

\begin{document}
\vspace*{1cm}
\title{Massive Stars and The Creation of our Galactic Center}
 \author{Donald F. Figer}
\affil{STScI, 3700 San Martin Drive, Baltimore, MD 21218 \\
JHU, 34th and Charles Street, Baltimore, MD 21218}

\begin{abstract}
Our Galactic Center hosts over 10\% of the known massive stars in the Galaxy.
The majority of these stars are located in three particularly
massive clusters that formed within the past 5~\Myr. While these
clusters are extraordinary, their formation repesents about half of the 
total inferred star formation rate in the Galactic Center. There is mounting
evidence that the clusters are just present-day examples of the
hundreds of such similar clusters that must have been created in the past, and
whose stars now comprise the bulk of all stars seen in the region. I
discuss the massive stellar content in the Galactic Center and present 
a new analysis that suggests that effects of continuous star formation in
the Galactic Center can be seen in the observed luminosity functions newly-obtained
HST/NICMOS and Gemini AO data. 
\end{abstract}

\section{Introduction}
Over 10\% of the known massive stars (M$_{\rm init}$$>$20~\Msun) in 
the Galaxy reside in three clusters of young stars
located within 30~pc of the Galactic Center. 
These clusters are the most massive young clusters in the Galaxy and 
contain approximately 30 Wolf-Rayet (WR) stars, at least 2 Luminous Blue
Variables (LBV), approximately a half dozen red supergiants, and approximately 300 O stars. 
Together, they emit enough ionizing radiation to account for roughly half of the thermal 
radio emission in the central few degrees of the Galaxy, suggesting
that the young clusters contain approximately half of the stars recently formed
in this region.
An additional collection of young stars exists
in the region, with members scattered about the central 50~pc; some have evolved
to the WR stage, while others are still deeply embedded within their natal
dust cacoons.

The current star formation rate can be approximated by dividing
the mass in newly formed stars by their ages, i.e.\ 5(10$^4$)~\Msun/5~\Myr$\sim$0.01~\Msun/yr,
or a star formation rate density of 10$^{-7}$~\Msun/yr~pc$^{-3}$. This rate is approximately
250 times higher than the mean rate in the Galaxy, and about the same factor lower than
the rate in starburst galaxies. 
Clearly, the Galactic Center has formed a plethora of stars in the past
5~\Myr, but it is less apparent when the millions of stars in the central
50~pc formed. If we assume that the star formation rate in the past was similar
to the present rate, then the total mass of stars formed over the past 10~\Gyr\ is
$\approx$10$^8$~\Msun\ within a radius of 30~pc of the Galactic Center, 
or an order of magnitude greater than this amount over the
whole Central Molecular Zone, as first suggested by  Serabyn \& Morris (1996). 

This review summarizes the current state of knowledge concerning the massive
stars in the Galactic Center, and the history of star formation therein.

\section{The Central Cluster}
The first young stars discovered in the GC are within
the central parsec of the Galaxy (Becklin \& Neugebauer 1968).
Rieke \& Lebofsky (1982) and Lebofsky, Rieke, \& Tokunaga (1983) confirmed the identification of blue and red supergiants in the
region, claiming that these stars formed in a burst of star formation $\sim$1~\Myr\ ago and
that their presence could account for the nearby ionized gas and heated dust.
Forrest et al.\ (1987) discovered a blue supergiant in the center having
a broad Brackett-$\alpha$ emission line, and Allen, Hyland, \& Hillier (1990) identified
this same star as having a spectrum similar to those for evolved massive stars in the Magellanic Clouds.
Further studies 
discovered that many of the blue supergiants are evolved massive stars with spectra having prominent
\HeI\ emission lines, firmly establishing a starburst event roughly 5~\Myr\ ago 
(Krabbe et al.\ 1991; 1995, Allen 1994, Rieke \& Rieke 1994, Blum et al.\ 1995, 
Eckart et al.\ 1995, Genzel et al.\ 1996, Tamblyn et al.\ 1996).

We now know that the Central Cluster contains over 30 evolved massive stars having \Minit~$>$20~\Msun. A current
estimate of the young population includes 9 WR stars, 20 stars with Ofpe/WN9-like 
{\it K}-band spectra, several red supergiants, and
many luminous mid-infrared sources in a region of 1.6~pc in diameter centered on Sgr~A$^*$ (Genzel et al.\
1996). In addition, I estimate that it contains 100 O-stars (O7 and later) still on the main sequence. 
Najarro et al.\ (1994) modeled the infrared spectrum of the ``AF'' star finding that it is
a helium-rich blue supergigant/Wolf-Rayet star, characterized by a strong
stellar wind and a moderate amount of Lyman ionizing photons. 
Najarro et al.\ (1997) expanded this work by analyzing spectra of 8 blue supergiants in
the center, finding extremely strong stellar winds ($\Mdot\sim$ 5 to 80 $\times 10^{-5}\, \Msunyr$),
relatively small outflow velocities (\Vinf $\sim$ 300 to 1,000~\kms), 
effective temperatures from 17,000\,K to 30,000\,K, stellar luminosities of 1 to $30 \times 10^{5}$ \Lsun,
and spectral characteristics consistent with an ``Ofpe/WN9'' classification. They concluded 
that the \HeI\ emission line stars  power the central parsec and belong to a young stellar cluster of massive stars
which formed a few million years ago. 
More recently, Paumard et al.\ (2001) reviewed the emission-line stellar population in the central parsec, using new
narrow-band infrared imaging. They found that the brightest emission-line stars divide into two categories,
bright narrow-line (200~\kms) and faint broad-line (1000~\kms) stars, the former being clustered
within a few arcseconds of Sgr~A*, and the latter being distributed between 5$\arcsec$ and 10$\arcsec$ from
the center.

Eckart et al.\ (1999) and Figer et al.\ (2000) identified massives stars 
within a few AU of the supermassive black hole. Evidently, a significant fraction of this small group
of stars are young ($\tau_{\rm age}<20~\Myr$) and require extraordinarily dense pre-collapse 
cores ($\rho>10^{11}~cm^{-3}$) or formation sites much further away from the central black hole
than their present location would suggest. Early results from proper-motion studies suggest that
at least some of the stars in this cluster are bound to the black hole and are not on highly
elliptical orbits (Ghez et al.\ 2001; Eckart et al.\ 2002); therefore they are likely to be near
to their formation sites.

\section{The Quintuplet Cluster}
	The Quintuplet Cluster is located approximately
30~pc, in projection, to the northeast of the Galactic Center  
(Glass, Catchpole, \& Whitelock 1987). In addition to the five 
bright stars for which the Quintuplet was named (Nagata et al.\ 1990;
Okuda et al.\ 1990), the Quintuplet cluster contains a variety of massive stars, including four WN,
five WC (possibly ten, see below), two WN9/Ofpe, two LBV, one red supergiant and several dozen less-evolved blue
supergiants (Figer et al.\ 1999a, 1999c). 
The five Quintuplet-proper members are massive stars (L~$\sim$~10$^5$~\Lsun) embedded within 
dusty cacoons, although their spectral types and evolutionary status are unknown (Moneti et al.\ 2001).
Figer et al.\ (1996, 1999a) argue that these stars are dust-enshrouded WCL stars, similar to WR 140 (Monnier et al.\ 2002)
and WR 98A (Monnier et al.\ 1999). In addition to these post-main sequence stars,
it is likely that 100 O-stars still on the main sequence exist in the cluster, assuming
a flat to Salpeter IMF.
The total cluster mass is estimated to be $\sim$10$^4$~\Msun; of this, the stars
with certain spectral identifications, i.e.\ the most massive ones, contribute
a few 10$^3$~\Msun. Given the extended distribution of the cluster, the 
implied mass density is greater than few thousand \Msun~pc$^{-3}$. 
The total ionizing flux is $\sim$10$^{51}$ photons~s$^{-1}$, enough 
to ionize the nearby ``Sickle'' HII region (G0.18$-$0.04). 
The total luminosity from the massive cluster stars is $\approx$ 10$^{7.5}$ \Lsun,
enough to account for the heating of the nearby molecular cloud, M0.20$-$0.033. 

The two LBVs are added to the list of 6
LBVs in the Galaxy. They include the Pistol Star (Moneti et al.\ 1994; Figer et al.\ 1995a, 1995b, 1998, 1999b; 
Cotera 1995; Cotera et al.\ 1996), one of the most luminous stars known,
and a newly identified LBV (Geballe et al.\ 2000) that is nearly as luminous as the Pistol Star. 
Both stars are luminous, ``blue,'' and variable, and the Pistol Star has ejected 10~\Msun\ of
material in the past $\sim$10$^4$~{\it yrs}, as evidenced by the remarkable Pistol nebula of ionized gas
surrounding the star (Figer et al.\ 1999b; Moneti et al.\ 2001).
Most of the luminous stars in the cluster are thought to be 3$-$5 Myr old, but significant age differences remain, i.e.,
the Pistol Star is thought to be $\approx$2~\Myr\ old.

\begin{figure}[t]
\caption{Mass function in the Arches Cluster. Note the absence of stars more massive than \Minit$>$300~\Msun,
implying the detection of an upper mass cutoff. This figure can be retrieved from 
http://nemesis.stsci.edu/\~figer/iau212\_imfcutoff.eps.}
\end{figure}

\begin{figure}[t]
\plottwo{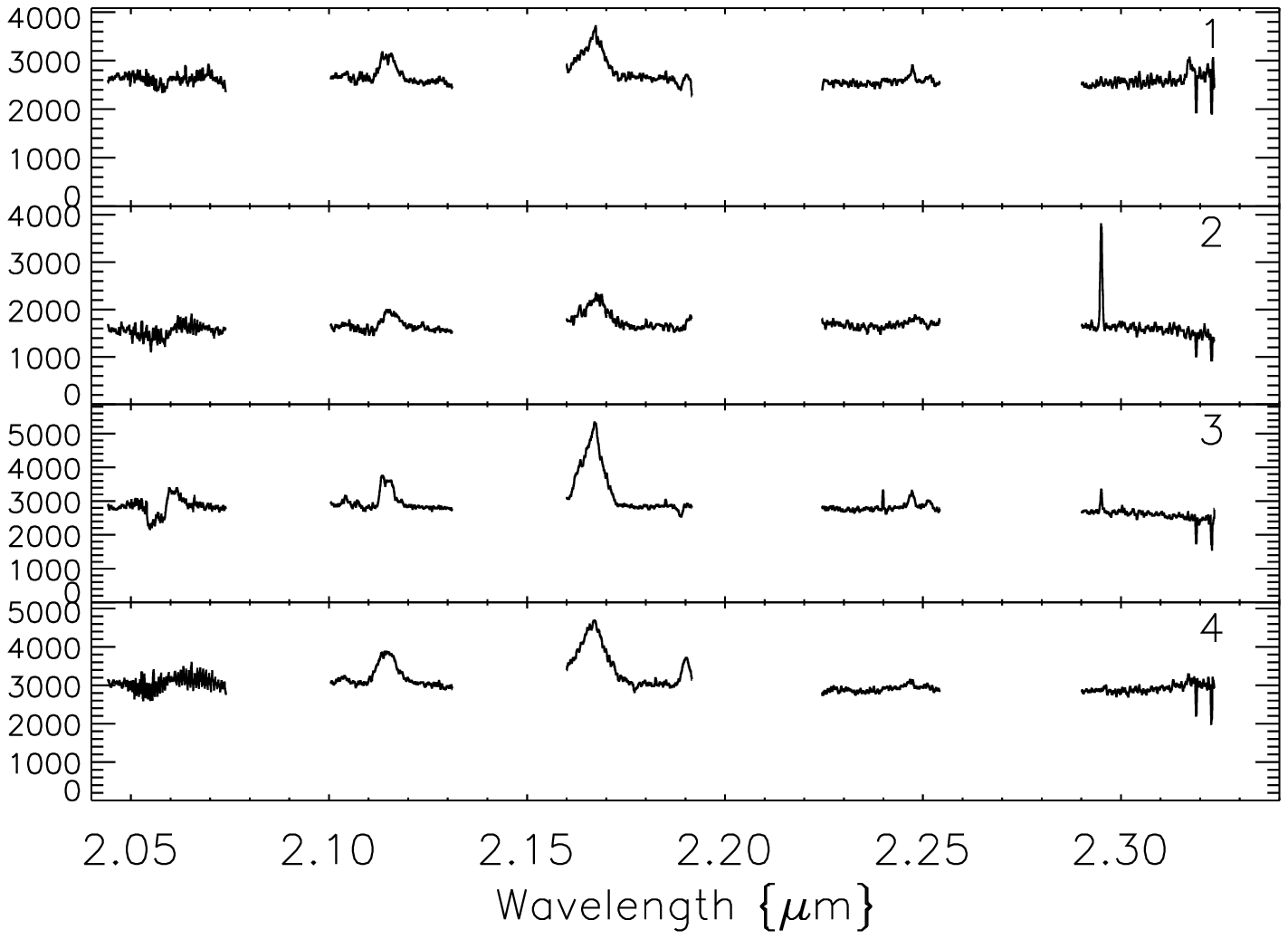}{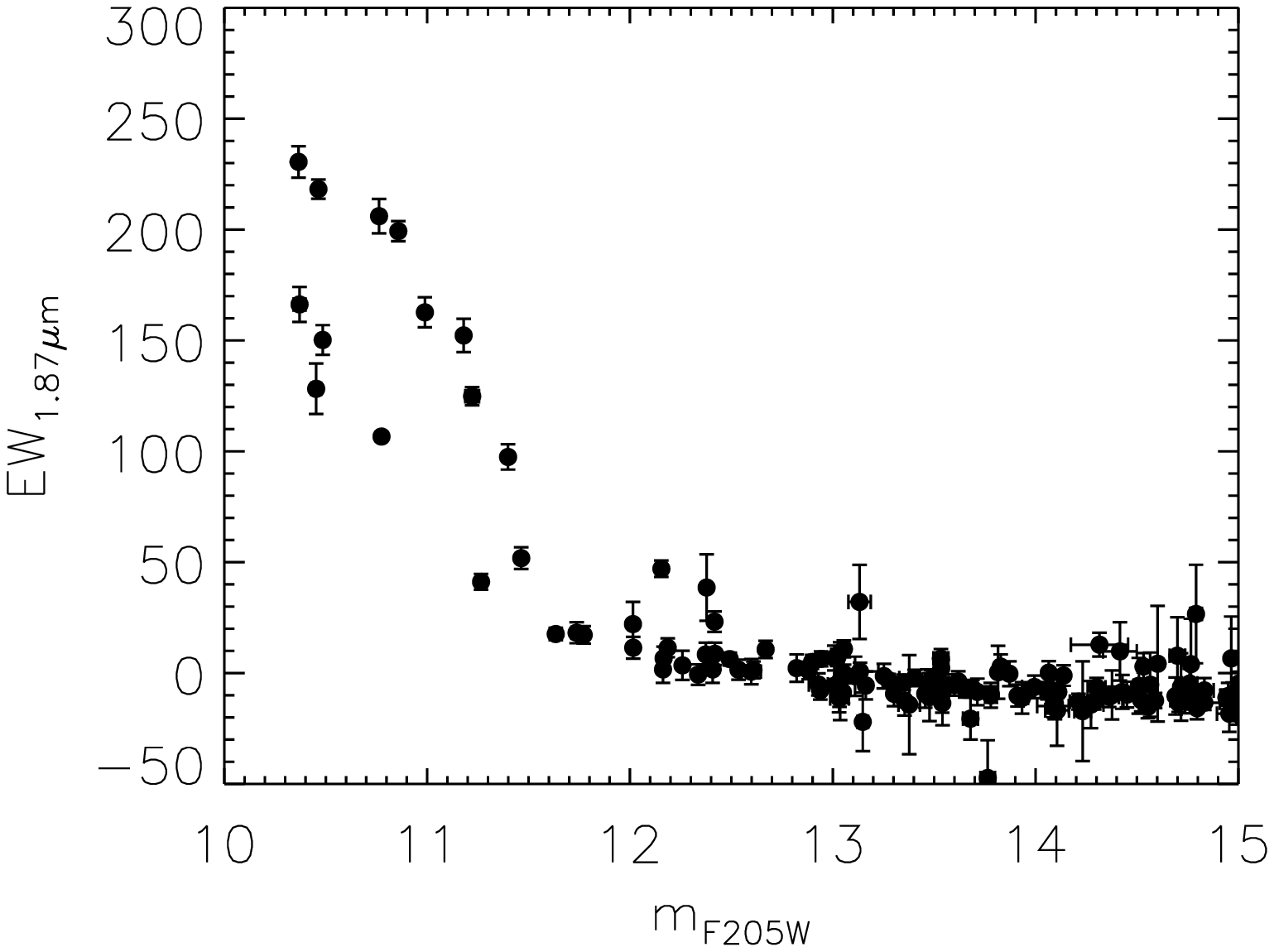}
\caption{{\it (left)} {\it K}-band spectra of four Wolf-Rayet stars (WNL) in the Arches Cluster from Figer et al.\ (2002).
Emission lines can be seen at 2.058~\micron\ (HeI), 2.104~\micron\ (NIII), 2.112/113~\micron\ (HeI),
2.115~\micron\ (NIII), 2.166~\micron\ (HeI/HI), 2.189~\micron\ (HeII), and 2.224/225~\micron\ (NIII).
The sharp feature near 2.32~\micron\ in the spectrum for star \#2 is due to a detector defect, and
the absorption features longward of 2.33~\micron\ are due to imperfect correction for telluric
absorption. {\it (right)} Equivalent width of the 1.87~\micron\ feature in massive stars of the Arches 
Cluster (Figer et al.\ 2002; see also Blum et al.\ 2001) as a function of apparent magnitude in the HST/NICMOS F205W filter.
The feature has contributions from HeI and HI.}
\end{figure}

\begin{figure}[t]
\centerline{\epsfxsize=9cm\epsfbox{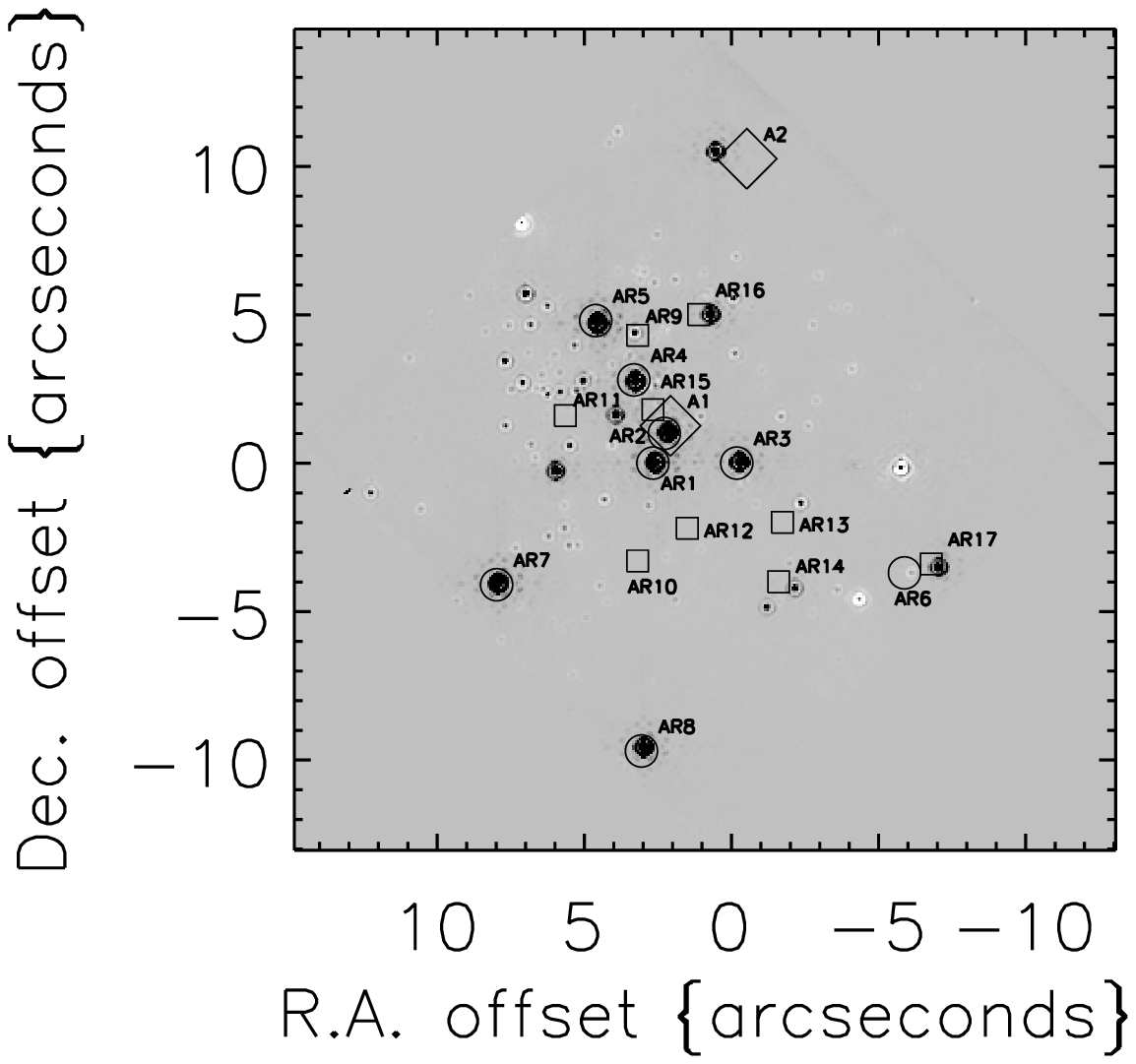}}
\caption{Difference image, F187N$-$F190N, highlighting stars with excess emission at 1.87~\micron.
Radio sources ($\nu=4.9~GHz$) are shown by circles, as identified in Lang et al.\ (2001), and
as squares, as identified by Figer et al.\ (2002). X-ray sources are shown by diamonds, as
identified by Yusef-Zadeh et al.\ (2002).}
\end{figure}

\section{The Arches Cluster}
First discovered about 10 years ago as a compact collection of a 
dozen or so emission-line stars (Cotera et al.\ 1992; Nagata et al.\ 1995; Figer 1995a; Cotera 1995; Cotera et al.\ 1996; Blum et al.\ 2001), 
the Arches cluster contains thousands of stars, including at least 160 O stars, according to Figer et al.\ (1999c). 
Figer et al.\ (1999c) used HST/NICMOS observations to estimate a total cluster 
mass ($\simgr$10$^4$~\Msun) and radius (0.2~pc) to arrive at an average mass density of 
3(10$^5$)~\Msun~pc$^{-3}$ in stars, suggesting that the Arches cluster is the densest, 
and one of the most 
massive, young clusters in the Galaxy. They further used these data to estimate an initial 
mass function (IMF) which is relatively flat ($\Gamma$~$\sim-$0.6$\pm$0.1) with respect to what has been 
found for the solar neighborhood ($\Gamma$~$\sim-$1.35, Salpeter 1955)
and other Galactic clusters (Scalo 1998). Stolte et al.\ (2002) recently confirmed this
flat slope by analyzing the same data and recently obtained Gemini AO data.
Figer et al.\ (2002) estimated
an age of 2.5$\pm$0.5~\Myr, based on the magnitudes, colors, mix of spectral types, and quantitative
spectral analysis of stars in the cluster. Given the current state of knowledge about
this cluster, it now seems apparent that we have observed a firm upper-mass cutoff, as
shown in Figure 1. Note that we should expect at least 10 stars more massive than \Minit=300~\Msun. 
Indeed, we should even expect one
star with an initial mass of 1,000~\Msun! Of course, it is questionable how long such a
star would live; however, it is clear that the Arches cluster IMF cuts off at around 150~\Msun.
Finally, even if we steepen the IMF slope to the Salpeter value, we still should expect at
least 4 stars more massive than 300~\Msun.

Figer et al.\ (2002) conclude that the most massive stars are bona-fide Wolf-Rayet (WR) stars and are some of the most massive stars known,
having \Minit~$>$100~\Msun, and prodigious winds, \Mdot~$>$10$^{-5}$~\Msunyr, that are 
enriched with helium and nitrogen. These findings are largely based upon the
spectra and narrow-band equivalent widths shown in Figure~2, and a detailed
quantitative analysis of these data (also see Najarro 2002).
Figer et al.\ (2002) found an upper limit to the velocity dispersion of 22~\kms, implying an
upper limit to the cluster mass of 7(10$^4$)~\Msun\ within a radius of 0.23~pc, and a 
bulk velocity of v$_{\rm cluster}\approx$+55~\kms\ for the cluster. 
It appears that the cluster happens to be ionizing, and approaching, the surface of a background 
molecular cloud, thus producing the Thermal Arched Filaments. They estimate that the cluster produces
4(10$^{51}$)~ionizing photons~s$^{-1}$, more than enough to account for the observed 
thermal radio flux from the nearby cloud.
Commensurately, it produces 10$^{7.8}$~\Lsun\ in total luminosity, providing the
heating source for the nearby molecular cloud, L$_{\rm cloud}\approx10^7$~\Lsun. 
These interactions between a cluster of hot
stars and a wayward molecular cloud are similar to those seen in the ``Quintuplet/Sickle''
region. Finally, note that significant work is being done on this cluster at radio
and x-ray wavelengths, i.e.\ shown in Figure 3.
 
\section{The Star Formation History of the Galactic Center}
The evidence for recent ($<$10~\Myr) star formation in the Galactic Center abounds.
The Lyman continuum flux emitted in the central few degrees of the Galaxy is
about 10$^{52}$ photons/s (Cox \& Laureijs 1989), with half coming from stars in the
three massive clusters. This flux is about 10\% of the that for the whole Galaxy, and 
the number of massive stars (\Minit~$>$20~\Msun) in the GC is about 10\% of the
number in the whole Galaxy. However, note that the star formation rate in the GC is about one
one-hundredth of that for the whole Galaxy. Such a low star formation rate as a 
function of Lyman continuum photon production necessarily
follows from the relatively flat initial mass function (IMF) slope used in estimating
the mass of stars formed in the young clusters. 

\begin{figure}
\plottwo{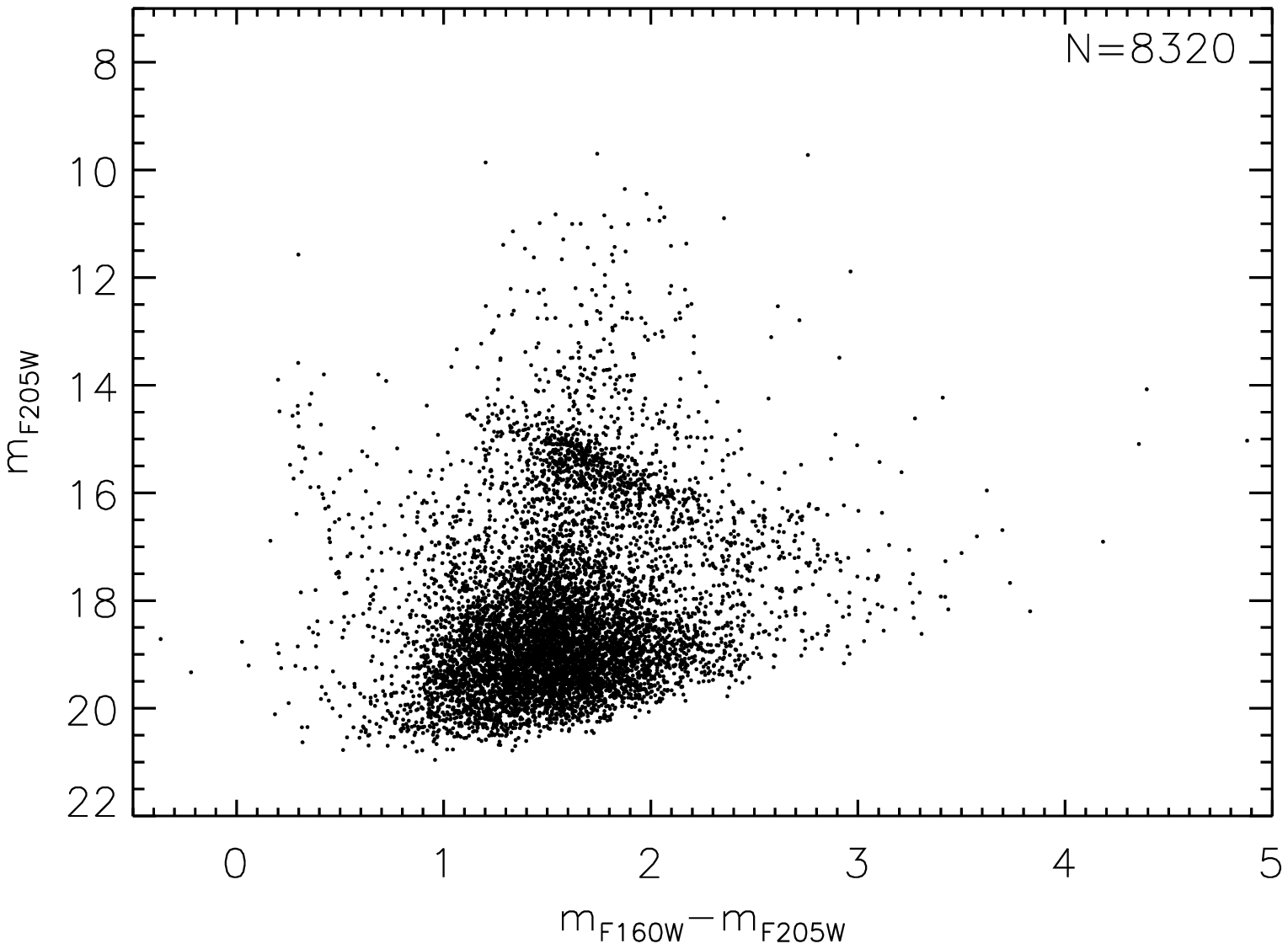}{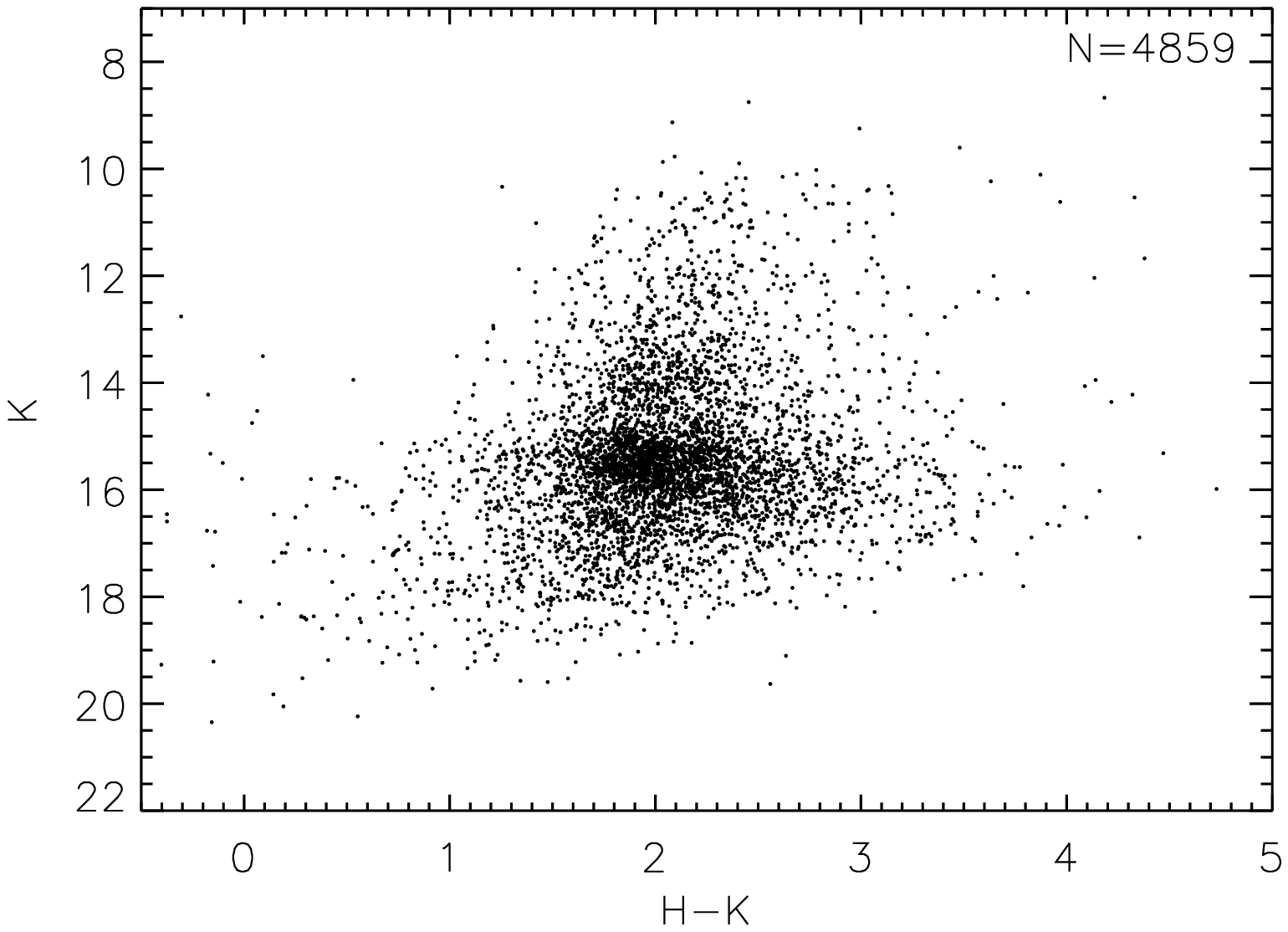}
\caption{{\it (left)} 
Infrared color-magnitude diagram for fields stars in the GC as obtained from NICMOS/HST data.
{\it (right)} Same, but from Gemini/AO data.}
\end{figure}

\begin{figure}
\centerline{\epsfxsize=12cm\epsfbox{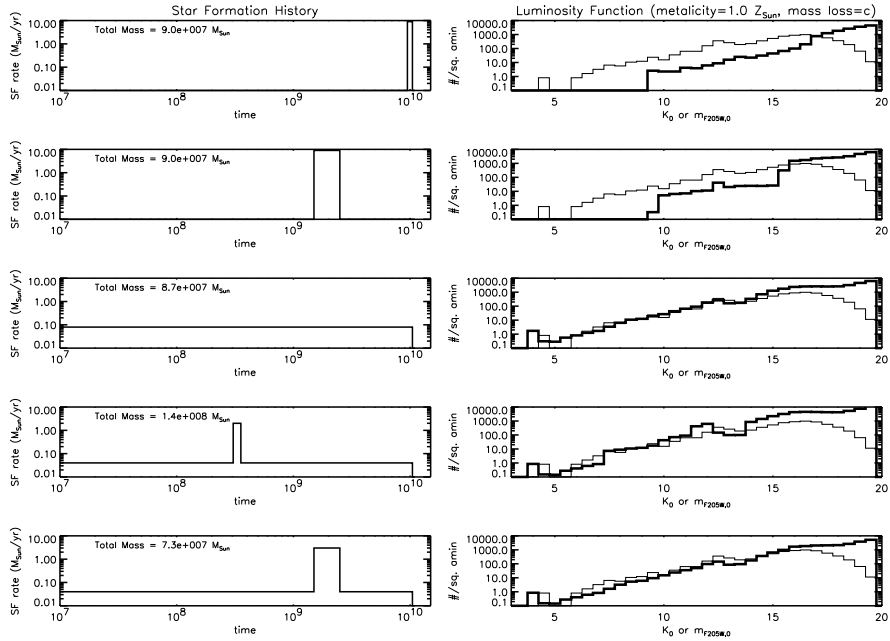}}
\caption{Model luminosity functions ({\it right, bold}) and observed luminosity functions ({\it right, light})
for various star formation histories in the Galactic Center ({\it left}). The models use the
Geneva isochrones with solar abundances and canonical mass loss rates.}
\end{figure}

The recent star formation history ($\tau_{\rm age}\simls50~\Myr$) in the Galactic Center
is relatively clear. Embedded HII regions trace star formation at the present time (Figer 1995,
Cotera 1995, Cotera et al.\ 1999), while the young clusters trace star formation that occured 2.5$-$5~\Myr\ ago. The lack
of red supergiants (M$_{\rm Bol}<-6.3$ and \Minit$>$8~\Msun) in the region provides evidence for a very low star formation rate from
5$-$20~\Myr\ ago. Indeed, a burst the size of that seen in the three clusters at 20 Myr ago 
would have produced 40 red supergiants, yet we see none (other than those associated with the young clusters).

The constraints on this activity are very strong, i.e.\ there were fewer than
5(10$^3$)~\Msun\ formed in stars over this time period, assuming the type of star
formation that spawned the three massive clusters. Looking beyond 20~\Myr,
the picture becomes less clear because it is difficult to separate old low-mass red giants from 
much younger high-mass AGB stars with photometric data alone. Using spectroscopy, Blum et al.\ (1996)
have shown that there is a relative dearth of stars with ages between 10 and 100~\Myr\ in the
central few parsecs. Given their conclusions, I estimate a low star formation rate during this period. 
These same authors, and Haller (1992), identified stars with ages on the order of a few hundred
\Myr\ in the central few arcminutes. Sjouwerman (1999) identified a population of OH/IR stars with
a narrow range of expansion velocities, indicating intermediate ages and a starburst event
a few \Gyr\ ago. In addition, Frogel et al.\ (1999) identified
an excess of bright stars in the fields they observed within 0$\fdg$2 of the Galactic Center.

In order to infer the past star formation rate, I modeled the observed 
surface number density of stars (Figure~4) as a function of star
formation history using the Geneva stellar evolution models, assuming a range of
power-law initial mass functions (IMFs), metallicity, and wind mass-loss rates. I considered: 
an ancient burst, episodic bursts, continuous formation,
and combinations of these three. I used these models to produce synthetic luminosity functions
for comparison with HST/NICMOS data. The surface number density in the observed luminosity functions 
has been set by dividing the number of stars per half magnitude bin by the area of the
observations. The NICMOS fields vary in location from about 15~pc from the
GC to 55~pc, and the sample was culled of forground and background stars by limiting
the data to stars with 1.0$<$\mnh$-$\mnk$<$3.0 (1.7$<$\AK$<$6.5). 
The results are presented in Figure~5.

In general, I find that the presently observed luminosity
function is very well fit by continuous star formation at the level of 0.05 to 0.1~\Msunyr. 
The total mass of stars formed is $\sim$10$^8$~\Msun\ in these models, although
the present-day mass is less as a result of mass-loss via steady winds or supernovae. 

An ancient burst model is not consistent with the presence of bright stars ({\it K}$_0$$<$8.0), nor is it consistent with
an enhanced brightness of the red clump (although the observational evidence for such an
enhancement is controversial). Episodic bursts are essentially indistinguishable from 
continuous star formation when using the luminosity function as a metric; clearly, one needs
to examine a subset of all stars in distinguishing burst widths and periodicities. The
continuous star formation model reproduces the bright end of the luminosity function, 
although note that the bulk of the observed bright stars are not red supergiants, so that
there must be a gap in significant star formation activity during recent times, excepting the
last 5~\Myr. 
I find that variations in metallicity, IMF slope, or mass-loss rate, do not qualitatively
affect our conclusion that continuous star formation produces synthetic luminosity functions
that best fit the observed luminosity function, compared with the other star formation
scenarios examined.

I acknowledge very useful discussions with Paco Najarro, Bob Blum, Laurant Sjouwerman, Mike Rich, Mark Morris, Sungsoo Kim,
and Jay Frogel.

\end{document}